\documentclass[conference]{IEEEtran}
\usepackage{cite}
\usepackage{amsmath,amssymb,amsfonts}
\usepackage{algorithmic}
\usepackage{graphicx}
\usepackage{textcomp}
\usepackage{xcolor}

\usepackage{hyperref}
\begin{document}

\title{Graph Transformation and Specialized Code Generation For Sparse Triangular Solve (SpTRSV)\\
}

\author{\IEEEauthorblockN{Buse Y{\i}lmaz}
\IEEEauthorblockA{\textit{Computer Science}\\
\textit{\.{I}stinye University}\\
\.{I}stanbul, Turkey \\
buse.yilmaz@istinye.edu.tr}
}

\maketitle

\begin{abstract}
Sparse Triangular Solve (SpTRSV) is an important computational kernel used in the solution of sparse linear algebra systems in many scientific and engineering applications. It is difficult to parallelize SpTRSV in today's architectures. The limited parallelism due to the dependencies between calculations and the irregular nature of the computations require an effective load balancing and synchronization mechanism approach. In this work, we present a novel graph transformation method where the equation representing a row is rewritten to break the dependencies. Using this approach, we propose a dependency graph transformation and code generation framework that increases the parallelism of the parts of a sparse matrix where it is scarce, reducing the need for synchronization points. In addition, the proposed framework generates specialized code for the transformed dependency graph on CPUs using domain-specific optimizations.
 \footnote{This paper is a revised version of the paper appeared in \url{http://www.basarim.org.tr/2020/doku.php} with the name "Buse Yilmaz and Didem Unat. \textit{Seyrek Alt Üçgen Matris Çözümü için Graf Dönüşümü ve Özelleşmiş Kod Üretimi}". The presentation in English can be found at (starts at min 10) \url{https://www.youtube.com/watch?v=fX2yM_gNHj8&list=PLRofGbhRbm7j3LaxVB3K_7hwOs1-1Serc&index=12&ab_channel=TUBITAKULAKBIMTV}}
\end{abstract} 


\section{Introduction}
From earth sciences to physics and chemistry, sparse linear systems are found in several applications of
science and engineering such as computational fluid dynamics, reservoir simulation and finite element
modeling. SpTRSV is the building block for several numerical solutions and it consumes a significant portion of the total execution time.  Direct and
iterative methods for sparse linear algebra systems and sparse matrix factorizations such as LU, QR and
Cholesky are examples to numerical solutions that use SpTRSV. Parallelizing SpTRSV has been studied
extensively and yet it remains a challenging problem even with today’s highly parallel and specialized
architectures, which deliver a tremendous amount of computation per second. The main challenges are: \textbf{1)} SpTRSV exhibits limited parallelism due to the dependencies between computations, \textbf{2)} Computations are irregular: workloads assigned to threads are fine-grained and vary in granularity due to the sparsity structure of the matrix, \textbf{3)} Challenges 1 and 2 result in significant synchronization and management overhead, \textbf{4)} Computational dependencies require a careful analysis and an efficient load balancing approach is needed due to the fine-grained workloads. 

The challenges arise from the nature of SpTRSV together with the sparsity pattern of the matrix:  the sparsity pattern is usually
non-uniform throughout the matrix. Therefore, a sparse matrix has parts exhibiting different degrees of
parallelism. SpTRSV cannot benefit from today’s highly parallel architectures for the parts that exhibit very few
parallelism. These parts usually have few rows, therefore, several cores sit idle due to the dependencies. 

To remedy the challenges mentioned above, we propose a dependency graph transformation
and code generation framework to:
\begin{itemize}
	\item Make the sparsity pattern of the matrix more uniform by increasing the parallelism where it is scarce
	\item Reduce the need for synchronization points
	\item Generate specialized code for SpTRSV on CPUs, using domain-specific optimizations 
\end{itemize}

To increase the degree of parallelism in parts of the matrix where the computation becomes serial, we
propose to alter the sparsity pattern by transforming the dependency graph by rewriting the equations rows represent. Rewriting the rows enables breaking the rows’ dependencies in a safe way, hence these rows can be moved to upper levels when level-set approach \cite{AndersonS89_Saad,Saltz:1990:AMS:3037529.3037535,ROTHBERG1992719} is used. Generating
specialized code for SpTRSV enables several optimizations such as reducing the memory accesses and indirect indexing, and applying arithmetic optimizations.

The rest of the paper is as follows: Section \ref{section:background} gives background information about SpTRSV and Section \ref{section:rewriting} introduces the equation rewriting approach. Matrix analysis and specialized code generation is presented in Section \ref{section:code_gen}, the following section, Section \ref{section:results}, presents the experiment results and we conclude in Section \ref{section:conclusion}.

\section{Background}
\label{section:background}
The sparse triangular solve takes a lower triangular matrix $L$ and a right-hand side vector $b$ and, it solves the
linear equation $Lx = b$ for $x$.

Figure 1 presents the operation $Lx = b$ together with the dependency graph of the lower triangular matrix $L$ partitioned into levels on the right. Dependency graph of $L$ is a directed acyclic graph (DAG\textsubscript{L}), where the nodes represent the rows and the edges represent the dependencies between nodes. Algorithm 1 on the right handside presents a serial forward substitution implementation of SpTRSV for solving \textit{Lz = b}.  \textit{n} represents the number of rows in \textit{L}. For each row, a partial sum  is calculated using the nonzeros (dependencies) in that row in the inner for loop. The inner loop can be parallelized for each row and the outer for loop can be parallelized to the extend that dependencies permit. Unless all dependencies are met, a row cannot be calculated. This causes a bottleneck on the parallelization of outer for-loop.

\begin{figure*}[h]
	\begin{minipage}{0.50\textwidth}
		\centering%
		\includegraphics[width=0.95\linewidth]{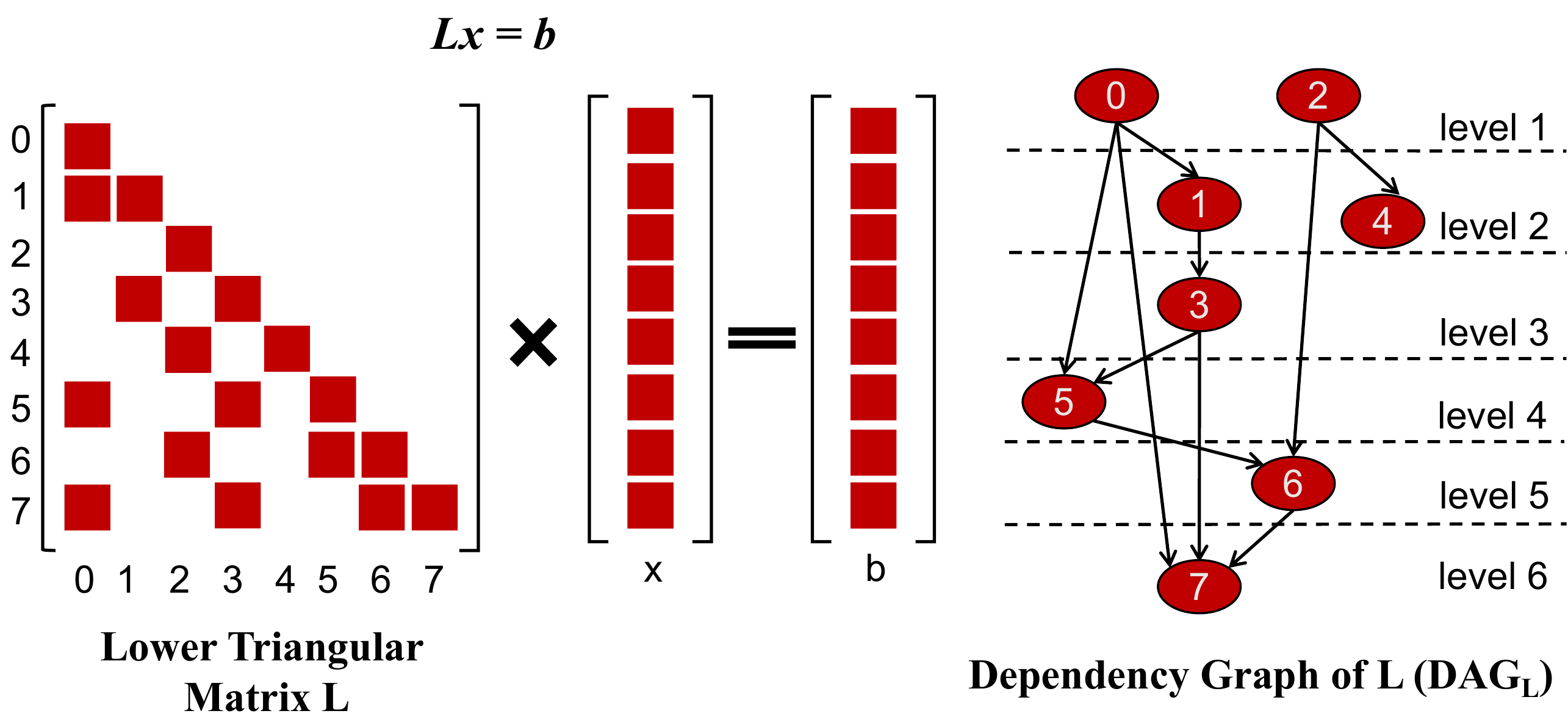}
	\end{minipage}
	\begin{minipage}{0.49\textwidth}
		\centering%
		\includegraphics[width=0.95\linewidth]{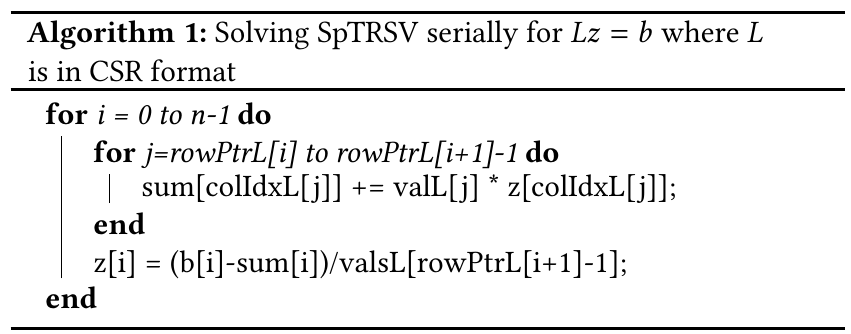}
	\end{minipage}
	\caption{\textbf{Left:} Lower triangular matrix $L$ and $L$'s dependency graph partitioned into levels \textbf{Right:} Serial algorithm for SpTRSV where the sparse matrix is in CSR format.}
	\label{fig:sptrsv_dag}
\end{figure*}

The literature is rich in approaches to optimize SpTRSV: Level-set methods \cite{AndersonS89_Saad,Saltz:1990:AMS:3037529.3037535,ROTHBERG1992719} group the rows that have no dependencies to each other into levels or wavefronts . The rows in a level are executed in parallel and the levels are executed one after the other (serially). In \cite{ROTHBERG1992719}, a level-set based method for SpTRSV is presented for CPUs where matrix reordering is used for the coefficient matrix to address the performance problem caused by poor spatial locality of the data.  Authors advocate for dynamic scheduling performing worse due to the fine-grained nature of the computations in SpTRSV. Later, this approach is adopted for NVIDIA GPUs in \cite{naumov_2011} and in \cite{Li2013}. A recent work extended the level scheduling for Sunway architecture \cite{Wang:2018:SFS:3200691.3178513}. The authors propose a new data layout, named sparse level tile (SLT) format, which improves data locality. 

Synchronization-free methods emerged as an alternative to level-set methods to eliminate the need for building level sets, and hence the need for barriers at the end of each level. First examples of this approach were on CPUs in \cite{hammond_schreiber_1992}. Later, several implementations on GPUs \cite{ALIAGA201979,Li_2017,liu_2017_fast_synchronization_free_algorithms,Liu:2016:SAP:2990973.2990990} have been proposed. In \cite{Li_2017}, authors use
a parallel topological sorting algorithm to set the levels and use a
counter-based scheduling mechanism, where each element only
waits for its own dependencies. In \cite{Liu:2016:SAP:2990973.2990990}, authors propose a simple
preprocessing phase, where self-scheduling mechanism is set up
based on the in-degree of dependency graph nodes.

The drawbacks of level-set and self-scheduling approaches are as follows:
\begin{itemize}
	\item A level consists of rows that can be run in parallel and a synchronization barrier at the end of each level is
required, causing the levels to be computed serially in a sequence. Therefore, using barriers incurs
	high synchronization overhead, especially for matrices with large number of levels. Computation of
	rows in a successor level has to wait for the current level to reach the barrier even if all dependencies of a
	row in a successor level are satisfied.
	\item Self-scheduling algorithms partition the rows into fine-grained tasks and a task is launched when its input
data is ready. While this method eliminates the need for synchronization barriers between
	consecutive levels, its implementation typically requires threads to busy-wait on their predecessor.
\end{itemize}

In \cite{Park:2014:SSH:2769884.2769893}, both level set and synchronization-free methods are utilized in a hybrid approach is developed for CPUs. In our previous study \cite{Buse_2020_adaptive_level_binning}, we adopted a similar approach to \cite{Park:2014:SSH:2769884.2769893}. Both works  aim to reduce synchronization points for SpTRSV on CPUs, eliminating unnecessary dependencies.

Another notable body of work about optimizing SpTRSV is done
in compiler domain. In \cite{sparso_2016},
authors enable context-driven optimizations such as matrix reordering by examining the properties of the sparse matrix and generate
code specialized for the matrix. Sympiler \cite{Cheshmi:2017:Sympiler}, analyzes the sparse codes symbolically during the compilation process and generates domain-specific code for various sparse matrix operations. In addition to the domain-specific optimizations similar to Sympiler, the specialized code generator proposed in this paper aims to reduce memory accesses by embedding the memory accesses into the code as constants.

Graph coloring is an NP-Complete problem that requires additional pre-processing and it is used in the optimization of SpTRSV.
Authors proposed algebraic block multicolor ordering in \cite{iwashita_2012} for
CPUs. Authors of \cite{Suchoski_2012}  and \cite{naumov_2015} apply graph coloring to
SpTRSV on GPUs. Block-diagonal based methods form another approach to improve SpTRSV. It is a method practiced on CPUs more than compared to GPUs. Matrix is reordered to help forming blocks around
the diagonal and off-diagonal to improve the locality \cite{Mayer2009,Smith:2011:STS:2076556.2076558,TOTONI2014454}.
In \cite{TOTONI2014454}, authors use dense BLAS operations to compute dense off-diagonal blocks. Dense matrix computations have higher parallelism than their sparse counterparts, therefore, creating dense blocks in sparse matrices is an effective method to improve performance.
These techniques are highly dependent on the sparsity structure of
the matrix.

\section{The Equation Rewriting Method}
\label{section:rewriting}
\begin{figure*}[!h]
	\centering%
	\includegraphics[width=0.90\linewidth]{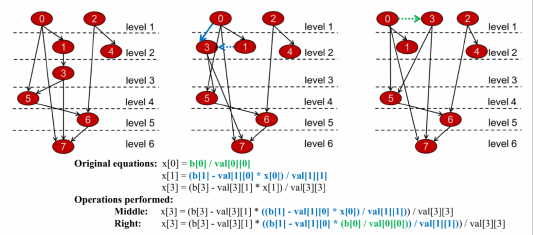}
\caption{\textbf{Left:} original equations before equation rewriting is applied \textbf{Middle:}  rewriting row $3$ using the original equation of row $1$ \textbf{Right:} rewriting row $3$ further, using the original equation of row $0$. The colored and dashed arrows indicate the broken bonds and the colored but steady arrows indicate the newly formed bonds.}
\label{fig:rewriting_operation}
\end{figure*}

In this work, we take the level set approach and transform the dependency graph of the given matrix by shifting rows from lower to upper levels using the equation rewriting method. However, the method is also applicable to synchronization-free approach since it works directly on the dependency graph. Transformation can take place to apply load balancing on the tasks too. 

The dependency graph is transformed by rewriting the equations each row represents leading to the fattening
of the thin levels (levels with a few rows) and removing the thin levels completely, reducing the number of the
synchronization barriers. In return, the idle cores can be utilized in the fattened levels increasing the
parallelism. Rewritten rows are calculated which normally must be calculated way after (due to dependencies)
at the cost of increased floating point operations (FLOPS) and sometimes increased memory access.

Figure \ref{fig:rewriting_operation} demonstrates the
graph transformation process. In this example, equation rewriting has been applied to row $3$ ($x[3]$) twice. The original equations of row $0$ and row $1$ are colored in green and blue. The first rewriting operation is also blue since it  uses row $1$, and the second rewriting operation is green since it uses row $0$. The dependency graph on the left is the original dependency graph, while the ones in the middle and on the right show the rows after the equation rewriting operations.

In the original dependency graph, row $3$ has a dependency on row 1 which depends on row $0$. As shown in the middle, by
replacing row $1$’s with its equation within the calculation of row $3$, its dependency on row $1$ is broken and now it
depends on row $0$, and row $3$ is shifted up from level $3$ to level $2$. Now, level $3$ is empty. Repeating the operation as shown on the right, row $3$ is
shifted up to level $1$ and now does not depend on any row. Since level $3$ is now empty it can be removed.

An important point to be noted is that the new equation specified with the label Figure \ref{fig:rewriting_operation} \textbf{Right} is not in the $Lx = b$ format anymore. This results in a significant increase in the number of arithmetic operations, hence in FLOPS. More importantly, the equations cannot be represented with a lower triangular matrix anymore. This may lead to losing the opportunity to use  existing SpTRSV algorithm for this matrix. To remedy this problem, a reorganization of the rewritten equations will be added to the rewriting process in the future. The equation  specified in Figure \ref{fig:rewriting_operation}, \textbf{Right} is rearranged to be in $Lx = b$ format and it  is shown in Figure 3. As seen in this figure, values of b vector are updated and the number of FLOPs decreases.

\begin{figure*}[!h]
	\centering%
	\includegraphics[width=0.95\linewidth]{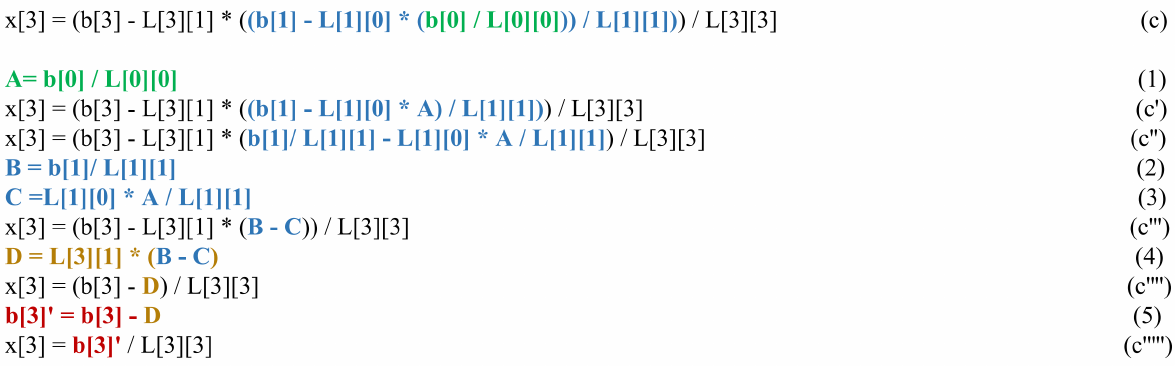}
	\label{fig:formatting}
	\caption{Rearrangement of the rewritten equation from Figure \ref{fig:rewriting_operation}, \textbf{Right}. Now the equation is in $Lx = b$ format. The $b$ vector entries are updated. Equations 1-5 are there for following the rearrangement easily.}
\end{figure*}

\section{Matrix Analysis and Code Generation}
\label{section:code_gen}
Matrix analysis module analyzes the input matrix and extracts matrix properties such as number of rows, number of nonzeros. When the analysis of the matrix is complete, DAG of the matrix and the level sets are constructed. Then additional information such as total number of memory accesses per level and average number of memory accesses per level is extracted from the DAG. This information is fed to the code generation module where equation rewriting process kicks in and code specialized for the matrix is generated. 

The specialized code generator is in very early stages of development. It produces code that will run in parallel by rewriting the equations of chosen rows and thus transforms the DAG accordingly. The code generator optimizes the code by reducing the memory accesses by converting them to constants and burying them into the code, and by eliminating the indirect indexing for rewritten rows. Several critical optimizations such as load balancing of level workloads, arithmetic optimizations and forming dense blocks to improve the locality are not implemented yet. An example to an arithmetic optimization is when the same memory accesses to vector x appear multiple times in the equation as a result of rewriting operation. These can be grouped together since each of them are only multiplied by different values. In addition, currently the rows to be rewritten are chosen manually by the programmer. The specialized code generator is written in C ++ and the code produced is C code with OpenMP \cite{openmp}  pragmas. 

Examples of the code generated are provided below in Figure 4 for matrix lung2 from SuiteSparse Matrix Collection \cite{suiteSparse}. On the left, the code generated when no equation rewriting is applied is shown. A function per level is generated by the specialized code generator until a threshold is hit: if the level is thick (having many rows), multiple functions are generated. Since equation rewriting is not applied, no rows are shifted between levels and thus no synchronization points are removed. Furthermore, memory accesses to vector x remain and only the others (to matrix $L$ and vector $b$) are converted to constants. On the right handside, the code generated when equation rewriting is applied is shown. Rows $2,3,4$ and $5$ which are at levels $1$ and $2$ are shifted to level $0$. Hence, the rows at higher levels depending on the rows $2,3,4$ and $5$ can be calculated immediately and the  levels $1$ and $2$ are removed together with their synchronization barriers at the end of these levels. 

\begin{figure*}[!h]
\centering%
\includegraphics[width=0.95\linewidth]{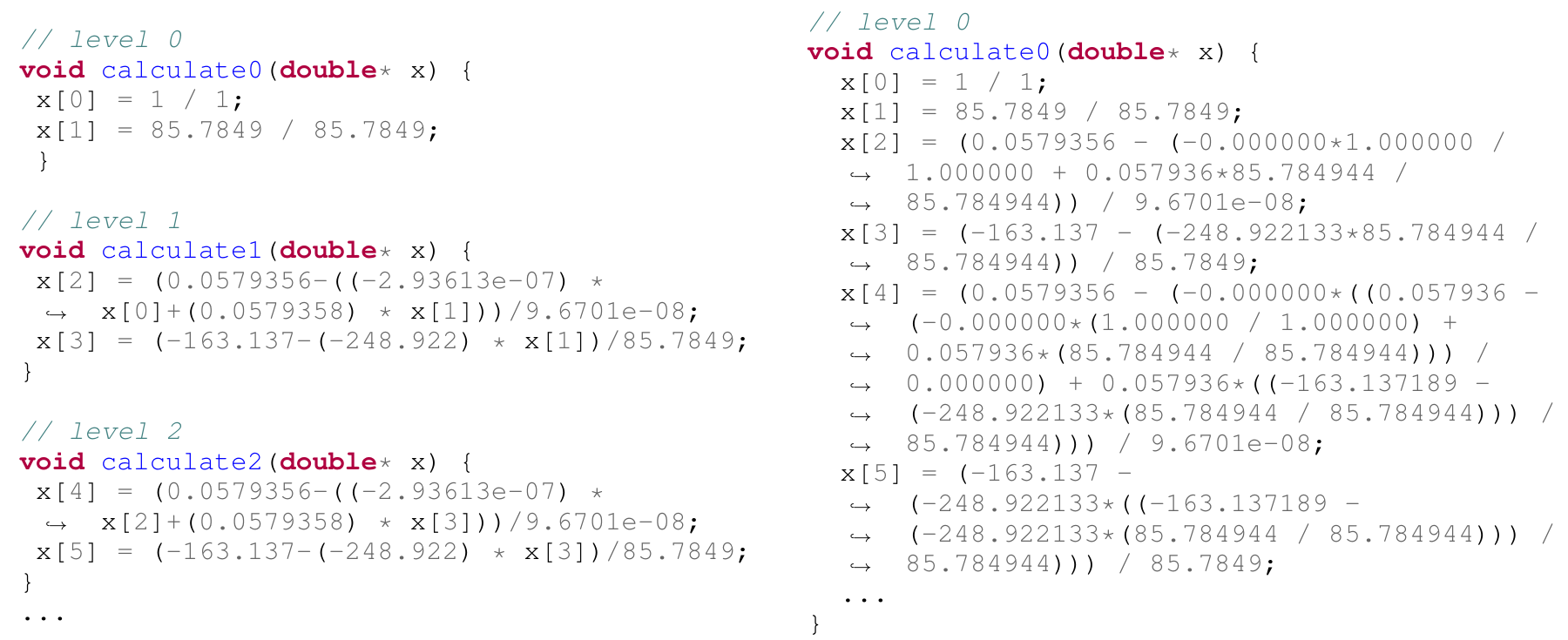}
\label{lst:lung2_rewritten}
\caption{\textbf{Left:}  Functions generated for lung2, when no rewriting is applied. Each function corresponds to a level in the original DAG  \textbf{Right:} Function generated for lung2 for the rows of level 0, when equation rewriting is applied. As a result of the equation rewriting process, rows of levels 1 and 2 are shifted to level 0.}
\end{figure*}

\begin{figure}[!h]
\centering%
\includegraphics[width=0.95\linewidth]{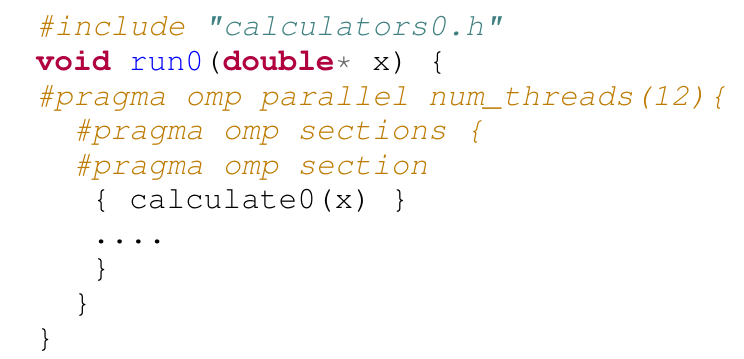}
\caption{Generated code that runs in parallel and calls the function shown on Figure \ref{lst:lung2_rewritten},  \textbf{Left}.}
\label{lst:lung2_driver}
\end{figure}

Figure \ref{lst:lung2_driver} shows the driver code generated by the specialized code generator. The function(s) generated for each level are called by openMP pragma directives for the parallel execution of the generated code. The pragma directives are not generated for a serial execution. All code is generated automatically along with a necessary spTRSV code and a makefile. Hence the executable is built with a single make command.

In both Figure \ref{lst:lung2_rewritten} and Figure \ref{lst:lung2_driver} very short or long functions are generated due to the lack of an efficient loaf balancing algorithm. 

\section{Experiment Results}
\label{section:results}
To observe the affects of the equation rewriting method on the dependency graph, we conducted two experiments. The matrix chosen for
this experiment is lung2 from SuiteSparse Matrix Collection \cite{suiteSparse}. lung2 has $109,460$ rows and $492,564$ nonzeros. After level sets are constructed, lung2 has $478$ levels (synchronization barriers) and 94\% of these levels have only $2$ rows, hence
they are very serial. 
The experiments are conducted on a dual socket Xeon Westmere with $2.53$GHz clock speed, 12
cores (24 threads) and 96GB DDR + 16 MCDRAM RAM. As the compiler, clang 5.0 \cite{10.5555/977395.977673} is used.

As mentioned in Section \ref{section:code_gen}, code generator is a prototype providing only reduction in memory accesses and elimination of indirect indexing and important optimizations such as load balancing of level workloads are not implemented yet.

In the first experiment, the performance of the specialized code generator is measured. No equation rewriting is applied and the code generated is serial. Average of $10$ iterations was measured as $1.98 ms$. As a comparison, a serial handwritten code using again the level set approach has an average of $10$ iterations as $1.14 ms$. An important note is that since lung2 is a very serial matrix, a serial implementation gives the better performance than its parallel counterparts. We observed that the performance of the prototype falls behind the compared serial code for the following reasons: \textbf{1)} the generated code is too long, \textbf{2)} there is no organization of the equations, same calculations are repeated, \textbf{3)} the code generator is written with generating parallel code in mind and there is no effective load balancing algorithm. Hence, there are unnecessary function calls. Functions can be very short and they should be merged. The aim of this experiment is to create a code skeleton to observe the benefits of the optimizations implemented as well as the optimizations planned. 

\begin{figure*}[!h]
	\centering%
	\includegraphics[width=0.95\linewidth]{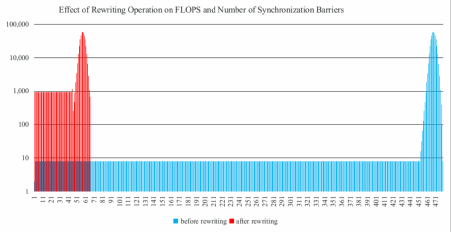}
	\label{fig:lung2_FLOPS}
	\caption{Number of FLOPS and number of levels before and after the equation rewriting method is applied. Rewriting rows reduces the number of synchronization barriers and increases the number of rows to
		be computed in a level}
\end{figure*}

In the second experiment the merits of the equation rewriting method is observed. As seen in Figure \ref{fig:lung2_FLOPS},  we applied equation rewriting to the thin levels (levels with very few rows) removing most of them completely by shifting
their rows to upper levels. After rewriting, $478$ levels dropped down to $66$, removing $86\%$ of the synchronization barriers. The number of total FLOPs to be performed increased only by $10\%$.  In this experiment which is carried out by running the generated code serially, the average of $10$ iterations was measured as $2.06 ms$. 

Considering that most of the proposed optimizations are not implemented, the lack of an effective load balancing method and equations are not reorganized, this experiment shows that \textbf{1)} the rewriting operation is promising in terms of removing the synchronization
barriers, \textbf{2)} matrices which deliver poor performance on parallel architectures and accelerators due to the parts
lacking parallelism, can now deliver better performance with fewer and fatter levels, \textbf{3)} benefit from the decrease in the
number of synchronization barriers and utilizing more cores is likely to surpass the performance loss due to
the increase in FLOPs, \textbf{4)} starting with an effective load balancing algorithm, optimizations mentioned in Section \ref{section:code_gen} are critical for performance gains.

\section{Conclusion}
\label{section:conclusion}
In this study, a new and original method of rewriting the equations of the rows of is proposed for the optimization of SpTRSV. We aim to increase the parallelism of the parts of a lower triangular matrix lacking parallelism by transforming them with the equation rewriting method. Thus, the dependency graph transformation allows the use of more cores by relaxing or completely eliminating row dependencies. In addition, thin levels can be completely eliminated by this method reducing the need for synchronization barriers. The equation rewriting method proposed is reinforced with a specialized code generator that can enable domain-specific optimization by generating code specialized to the matrix. Performance results show that the rewriting method is promising for performance optimization of SpTRSV.


\end{document}